# Capacity Optimized For Multicarrier OFDM-MIMO Antenna Systems

Nirmalendu Bikas Sinha, Prosenjit Kumar Sutradhar And M.Mitra

**Abstract**— Motivated by MIMO broad-band fading channel model, in this section we deals with the capacity behaviour of wireless MIMO and OFDM based spatial multiplexing systems in broad-band fading environments for the case where the channel is unknown at the transmitter and perfectly known at the receiver. This influence the propagation and system parameters on ergodic capacity, we furthermore demonstrate that, unlike the single-input single-output (SISO) case, delay spread channels may provide advantage over flat-fading channels not only in terms of outage capacity but also in terms of ergodic capacity. Therefore, MIMO delay spread channels will in general provide both higher diversity gain and higher multiplexing gain than MIMO flat-fading channels.

**Index Terms**— MIMO, OFDM ,ISI, CP.

——————————— ◆ ———————————

## 1. INTRODUCTION

Digital communication using MIMO processing has emerged as a breakthrough for revolutionary wireless systems. It solves two of the toughest problems facing any wireless technology today: speed and range. But this MIMO system having fast framing rate of the order of 1-2 □s will be polluted by ISI when operational in an environment having a typical time delay spread of 200 □s. Thus an ISI value of 200/2 =100 is an undesirable multi-path effect for the real MIMO system. Therefore MIMO cannot achieve zero ISI and hence cannot be utilized alone. A potential disadvantage of MIMO is its complexity, particularly the broadband time-domain version of MIMO. In the presence of delay-spread, a MIMO receiver (space-time equalizer) is quite complex. MIMOs can have potentially been combined with any modulation or multiple access techniques. Among the existing air-interface techniques, OFDM based multi-carrier approach may be enabler for the MIMO broadband operation So the fast frames are slowed down first and converted to several slow sub frames and modulated to multiple carriers of OFDM. The use of MIMO technology in combination with OFDM, i.e., MIMO-OFDM [1-3], therefore seems to be an attractive solution for future broadband wireless systems. Because in MIMO-OFDM, OFDM simplifies the implementation of MIMO without loss of capacity, reduces receiver complexity, avoids inter-symbol-interference by modulating narrow orthogonal carriers and each narrow-band carrier is treated as a separate MIMO system with zero delay-spread.

————————————————

- *Prof. Nirmalendu Bikas Sinha, corresponding author is with the Department of ECE and EIE , College of Engineering & Management, Kolaghat, K.T.P.P Township, Purba- Medinipur, 721171, W.B., India.*

- *Prosenjit kumar sutradhar is with the Department of ECE, College of Engineering & Management, Kolaghat, K.T.P.P Township, Purba- Medinipur, 721171, W.B., India.*

- *Dr. M.Mitra is With the Bengal Engineering and science University, Shibpur, Howrah, India .*

## 2. OFDM-MIMO (V-BLAST) SYSTEM MODEL

The multiuser MIMO-OFDM scheme which includes arrays of M transmit antennas performs a MIMO vertical encoding (VE)/convolution encoding and N receive antennas (V-BLAST requires N ≥ M) is illustrated in Fig. 1(a) and Fig.(b). Assume that the system is operating in a frequency-selective Rayleigh fading environment and that the channel coefficients remain constant during a packet transmission, i.e., quasi-static fading.

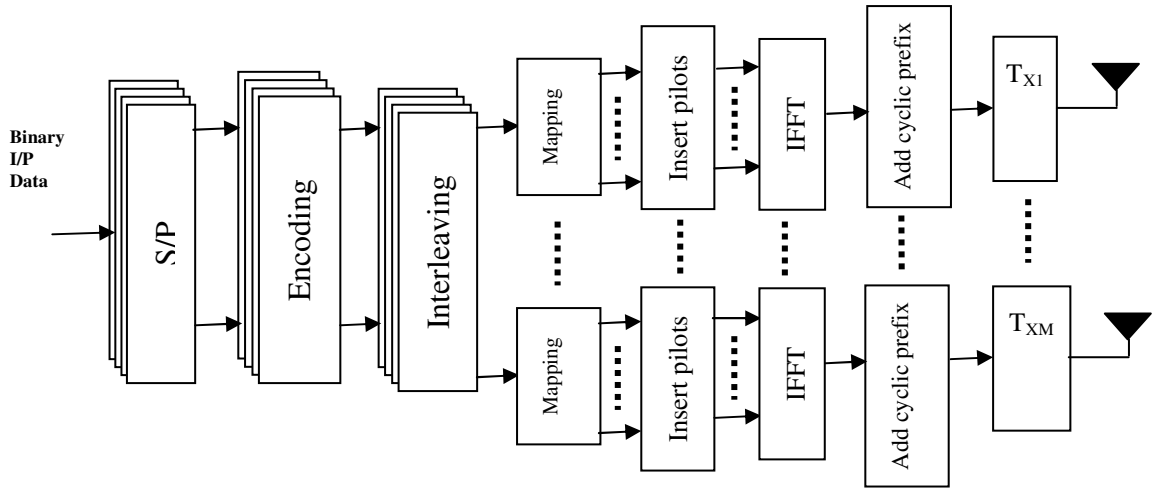

Fig. 1(a): Block diagram of OFDM-MIMO Transmitter System.

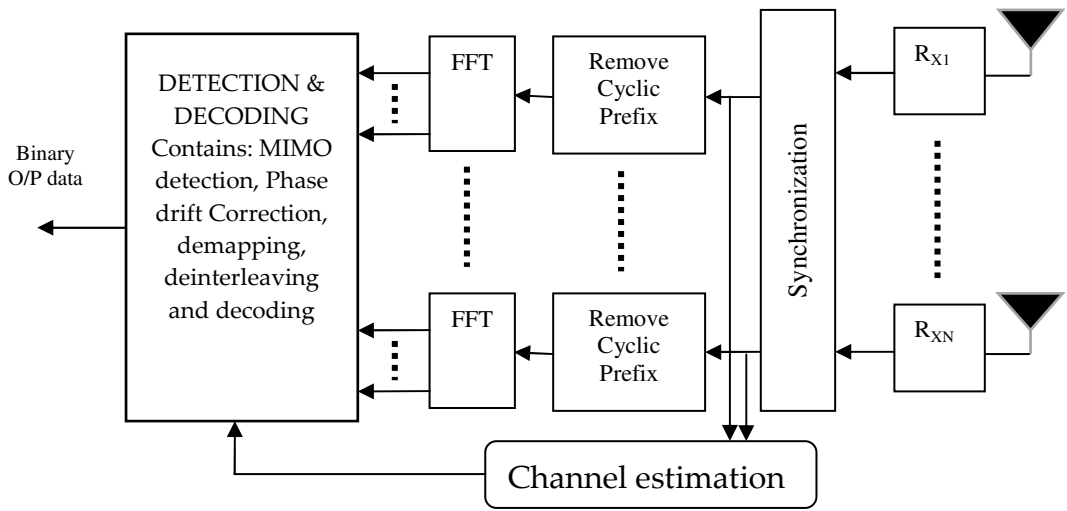

Fig.1(b): Block diagram of OFDM-MIMO Receiver System

The MIMO OFDM V-BLAST system operates in the 17 GHz unlicensed frequency band with an available bandwidth of 200 MHz (17.1–17.3 GHz) that is divided into four 50 MHz-width channels not simultaneously selectable. The binary input data is initially sent to a length 127 pseudorandom sequence scrambler (frequency sub channels) is designed for each of these 50 MHz wide channels. The purpose of the scrambler is to prevent a long sequence of 1s and 0s. The signal is then send a convolution/VE encoder. The de facto standard for this encoder is (2,1,7). The other rate ½ is achieved by puncturing the output if this encoder. Puncturing involves deleting coded bits from output data sequence, such that ratio of uncoded bits to coded bits is greater the mother code. The signal is then sent to an interleaver. The idea of interleaving is to disperse a block of data in frequency so that the entire block does experience deep fade in the channel. This prevents burst errors at the receiver. Otherwise the convolution /VE decoder will not perform very well in presence of burst errors. The interleaved are grouped together to form symbols. Therefore the coded bits are then mapped to some symbols. It has been established that OFDM is a spectrally efficient modulation technique, thus spectral efficiency depends mainly on the bandwidth of the symbol ($B_S$). This depends on the modulation technique used to modulate the individual subcarriers. It is the mapping (over a constellation) that corresponds to the choice of modulation technique which should minimize $B_S$. The symbols are then modulated using BPSK, QPSK and 16QAM schemes. These symbols are put through the top IFFT modulator and bottom to the lower IFFT modulator. Because each input to IFFT corresponds to an OFDM subcarrier, at the output we get a time-domain OFDM symbol that corresponds to the input symbols in the frequency domain. In other words, the symbols constitute the frequency spectrum of the OFDM symbol. These subcarriers are then padded with zeroes to make the full OFDM symbols, then convert this form parallel to serial in a multiplexer and append the cyclic prefix (CP), then transmitted by the antenna M. Once we have the OFDM symbol, a cyclic extension (with length depending on the channel) is performed. The final length of the extended OFDM signal will be the length of the original OFDM symbol plus the length of the channel response. As long as the guard interval, which is another name for the cyclic extension, is longer than the channel spread, the OFDM symbol will remain intact. Hence the information is transmitted in packets. The Rx is the exact inverse process after the incoming packets are stripped of their CP. The quantum of degradation depends on the type of constellation used and the spectral width of the phase noise of the carrier oscillator. For BPSK, there is no appreciable degradation, but in the case of 16 QAM this degradation is 0.5 dB for a phase noise of 40 KHz. Hence, the 10 Hz phase noise is considered negligible.

## 3. PERFORMANCE EVALUATION

*3.1 Ergodic Capacity of MIMO system*

The ergodic capacity of a MIMO channel is the ensemble average of the information rate over the distribution of the elements of the channel matrix H . It is the capacity of the channel when every channel matrix H is an independent realization. This implies that it is a result of infinitely long measurements. Since the process model is ergodic, this implies that the coding is performed over an infinitely long interval. Hence, it is the Shannon capacity of the channel. Ergodic capacity of a MIMO system can be expressed as

$$C_{\text{MIMO Ergodic Capacity}} = \in \left\{ \sum_{i=1}^{r} \log_2 \left( 1 + \frac{\varrho}{M} \lambda_i \right) \right\} \ldots \ldots (1)$$

Where r is the rank of the channel, $\varrho$ = signal to noise ratio ($E_s/N_o$) and $\lambda_i$ (i = 1,2,...r) are the positive eigen values of $HH^H$. The expectation operator applies in this case because the channel is random. Since H is random, the information rate associated with it is also random.
Fig.2 shows the ergodic capacity over different system configurations as a function of SNR. We note that ergodic capacity increases with increasing $\varrho$ and with increasing M and N.

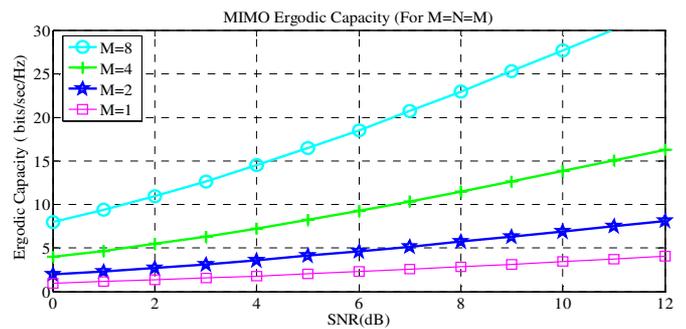

Fig.2 Ergodic capacity for different antenna configurations with M=N=M.

The ergodic capacity when the channel is known to the transmitter is always higher then when it is unknown. This advantage reduces at high SNRs. This is because at high SNRs, all eigen channels perform equally well (i.e., there is no difference in quality between them). Hence, all the channels will perform to their capacities, making both cases nearly identical.

### 3.2 *Outage Capacity or Nonergodic capacity for MIMO system*

Outage (or failure) capacity is the capacity that guaranteed with a certain level of reliability. In this case, capacity is viewed as a random entity [4], [5] since it depends on the instantaneous random channel parameters. This situation typically occurs when stringent delay constraints are imposed, as is the case, for example, in speech transmission over wireless channels. A capacity in the Shannon sense does not exist since, with nonzero probability, which is independent of the code length, the mutual information falls below any positive rate, as small as it may be. We define x % outage capacity as the information rate that is guaranteed for (100-x) % of the channel realizations, that is, $x (C \leq C_{outage}) = x$ % [6].

Fig. 3 shows the 10% outage capacity for several MIMO cases, when the channel is i.i.d. and unknown at the transmitter. We note that as the SNR increases, the capacity increases and as the number of antennas increases, so does the capacity. We note that for every 3 dB increase in SNR the capacity of MIMO increases M bits/s/Hz as compared with 1 bits/s/Hz in a SISO channel.

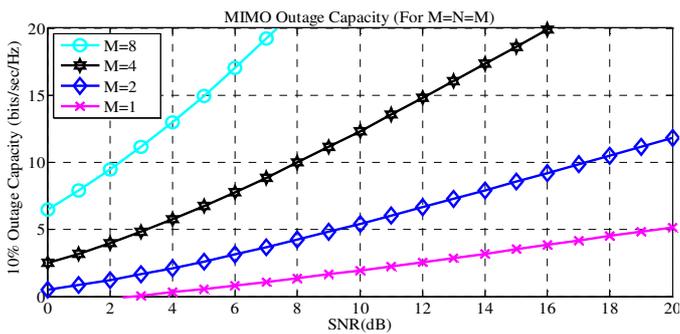

Fig.3 10% outage capacity for various antenna configurations.

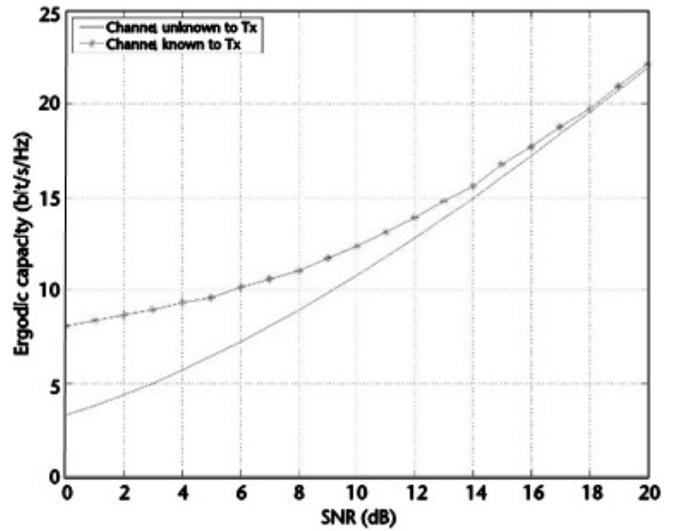

Fig.4 10% outage capacity of an M = 4 channel with and without channel knowledge at the transmitter.

The difference in outage capacities decreases with SNR. If the channel is known at the transmitter, Fig.4 shows that water-filling is a superior solution.

### 3.3. *Ergodic Capacity for MIMO-OFDM system:*

For a broad-band MIMO fading channel model, which is based on previous work reported in [7], [8], we provide expressions for the ergodic capacity and the outage capacity of OFDM-based spatial multiplexing systems [9], [10] considering the case where the channel is unknown at the transmitter and perfectly known at the receiver. These expressions are then used to study (analytically and numerically) the influence of propagation parameters such as delay spread, cluster angle spread, and total angle spread, and system parameters such as the number of antennas and antenna spacing on capacity. We find that, in the MIMO case, unlike the single-input single-output (SISO) case, delay spread channels may provide an advantage over flat-fading channels not only in terms of outage capacity but also in terms of ergodic capacity. Consequently, MIMO delay spread channels provide not only higher diversity gain than MIMO flat-fading channels but also higher multiplexing gain.

The use of OFDM in the context of spatial multiplexing has been proposed previously in [9], [10], [11], and [12]. However, it appears that no capacity studies of OFDM-based spatial multiplexing systems using the physically motivated MIMO fading channel model provided in this paper have been performed so far. For the single-carrier narrow-band flat-fading case, the impact of spatial fading correlation and antenna array geometry on capacity has been studied in [13] and [14]. To the best of our knowledge, the impact of physical parameters (delay spread, cluster angle spread, and total angle spread) and system parameters (number of antennas and antenna spacing) on ergodic capacity and outage capacity in the broad-band OFDM case has not been studied in the literature so far.

Fig. 1 shows a schematic of an OFDM-based spatial multiplexing system. In this section, we assume that the length of the cyclic prefix (CP) in the OFDM system is greater than the length of the discrete-time baseband channel impulse response. This assumption guarantees that the frequency-selective fading channel indeed decouples into a set of parallel frequency-flat fading channels [14].

Let consider $C_D^{(i)}$ is the data symbol transmitted from the $i^{th}$ antenna on the $D^{th}$ tone. Where $D^{th} = (0,1,......,N-1)$ tone,

$$H(e^{j2\pi\theta}) = \sum_{l=0}^{L-1} H_l\, e^{-j2\pi l\theta} \quad (0 \le \theta < 1)$$

and $C_D = \begin{bmatrix} c_D^{(0)} & c_D^{(1)} & \ldots & c_D^{(M-1)} \end{bmatrix}^T$ is the transmitted data symbols into frequency vectors. It can be shown that

$$\tilde{C}_D = H(e^{j2\pi(K/n)})C_D + n_D,\; D = 0,1\ldots n-1 \ldots\ldots (2).$$

Where $\tilde{C}_D$ denotes the reconstructed data vector for the $D^{th}$ tone and $n_D$ is additive white Gaussian noise (AWGN) satisfying $\varepsilon\{n_D n_l^H\} = \sigma_n^2 I_N \sigma[k-l] \ldots\ldots(3)$. From equation (2) it can be seen that equalization requires application of a narrowband receiver for each tone $D = 0,1,\ldots n-1$. Equation (2) can be rewritten as

$$\tilde{C} = H_c + n \ldots\ldots(4).$$

We assume that each of the OFDM symbols transmitted uses a independent realization of the random channel impulse response matrix and that the channel matrix remains constant within one channel use. Using equation 4, the mutual information (in b/s/Hz) of the OFDM based spatial multiplexing system under an average transmitter power constraint is given by [15], [16],

$$I = \frac{1}{N}\log_2\left[\det\left(I_{N,n} + \frac{1}{\sigma_n^2} H \sum H^H\right)\right] \ldots (5)$$

where $T_r(A)$ is the trace of the matrix A. $\sum$ with $T_r(\sum) \le P$ is the covariance matrix of the Gaussian input vector C and P is the maximum overall transmit power. Note that mutual information is normalized by n, since n data symbols are transmitted in one OFDM symbol and that we ignored the loss in spectral efficiency due to the presence of the CP. The nM×nN matrix $\sum$ is a block diagonal matrix given by

$$\sum = \text{diag}\{\sum_k\}_{D=0}^{n-1}.$$

Where the M×M matrices are the covariance matrices of the Gaussian vectors $C_D$, and as such determine the power allocation across the transmit antennas and across the OFDM tones. If the channel is perfectly known at the transmitter, the optimum power allocation is obtained by distributing the total available power P according to the water-filling solution [9]. In OFDM-based spatial multiplexing systems, statistically independent data symbols are transmitted from different antennas and different tones and the total available power is allocated uniformly across all space–frequency sub channels [9],[10]. Therefore, we set

$\sum_D = \left(P/(Mn)\right) I_M\; (D = 0,1,\ldots n-1)$, which is easily verified to result in $T_r(\sum) = P$ Using (equ. 5), we therefore obtain

$$I = \frac{1}{n}\sum_{D=0}^{n-1} I_D$$

$$I = \frac{1}{n}\sum_{D=0}^{n-1} \log_2\left[\det\left(I_N + \rho H(e^{j2\pi(D/n)})H^H(e^{j2\pi(D/n)})\right)\right] \ldots (6)$$

Where, $\varrho = P/(Mn\sigma_n^2)$. The quantity $I_D$ is the mutual information of the $D^{th}$ MIMO-OFDM sub channel. Since $H(e^{j2\pi(D/n)})$ is random, $I_D$ is a random. The distribution of $I_D(D = 0,1, n-1)$ is independent of D. Therefore,

$$I_D \sim \log[\det(I_N + \varrho \wedge H_w H_w^H)] \ldots (7),$$

Where, $\wedge = \text{diag}\{\lambda_i(R)\}_{i=0}^{N-1}$, $H_w$ is an M× N i,i,d random matrix. $R = \sum_{l=0}^{L-1} R_l$ and $\lambda_i(R)$ denotes the i$^{th}$ Eigen value of R. The Ergodic capacity for OFDM-MIMO system from equation (6)

$$(C_{\text{OFDM−MIMO Ergodic capacity}}) = \in \left\{\frac{1}{n}\sum_{D=0}^{n-1} I_D\right\} \ldots (8)$$

Utilizing equation (7) equation (8) reduces to

$$C = \left\{\log_2\left[\det(I_N + \varrho \wedge H_w H_w^H)\right]\right\} \ldots (9)$$

Case 1: For large M and fixed N

$$(1/M)H_w H_w^H \to I_N$$

$$C_{\text{OFDM−MIMO Ergodic capacity}} = \log_2[\det(I_N + \bar{\varrho} \wedge)] \ldots (10)$$

$$\bar{\varrho} = M\varrho = (P/N\sigma_n^2)$$

Case 2: For low SNR, i.e., for small $\bar{\rho}$ and large M

Equ.(10) reduces to

$$C_{\text{OFDM−MIMO Ergodic capacity}} = \log_2\left(\prod_{i=0}^{N-1}\left(1 + \bar{\varrho}\lambda_i(R)\right)\right) = \log_2[1 + \bar{\varrho}T_r(R)] \ldots (11)$$

Thus quantum delay spread in the channel has no impact on ergodic capacity when the SNR is low.

Case 3: For high SNR

$$C_{\text{OFDM−MIMO Ergodic capacity}} = \sum_{i=0}^{N-1}\log_2\left(1 + \bar{\rho}\lambda_i(R)\right) \ldots (12)$$

This implies that the eigenvalue spread of the sum correlation matrix $R = \sum_{l=0}^{L-1} R_l$, therefore critically determining the ergodic capacity(Fig.5). The more the eigenvalues, the more are the delay paths through the channel. The more the delay paths(Fig.6), the more routes are available from the transmitter to the receiver. The number of eigenvalues determines the rank of the sum correlation matrix. Each delay path contributes to one eigenvalue. Hence, the more the delay paths, the higher the capacity of the channel.

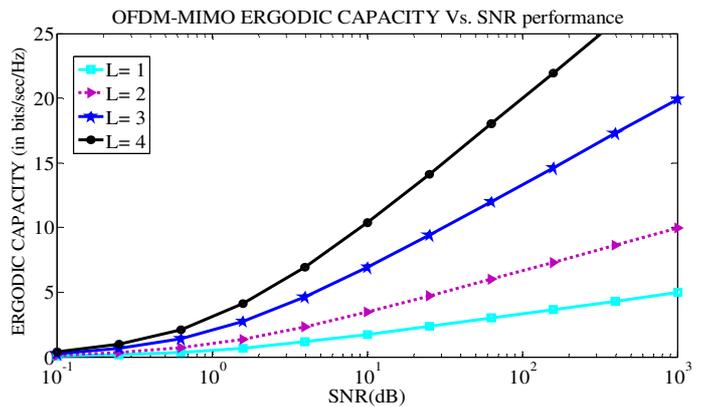

Fig. 5. Ergodic capacity (in b/s/Hz) as a function of SNR for various values of L

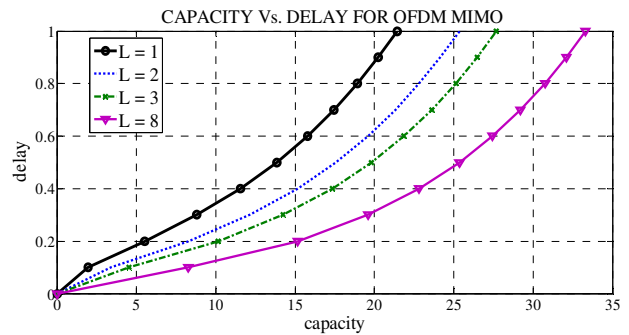

Fig. 6. Ergodic capacity (in b/s/Hz) as a function of delay factors for various values of L

## CONCLUSION

Based on a physically motivated model for broad-band MIMO fading channels, we derived expressions for the ergodic capacity and for outage capacity of OFDM-based spatial multiplexing systems for the case where the channel is unknown at the transmitter and perfectly known at the receiver. We studied the influence of propagation parameters and system parameters on ergodic capacity and outage probability and demonstrated the beneficial impact of delay spread and angle spread on capacity. Specifically, we showed that, in the MIMO case as opposed to the SISO case, delay-spread channels may provide advantage over flat-fading channels not only in terms of outage capacity but also in terms of ergodic capacity (provided the assumption that delayed paths tend to increase the total angle spread is true). Furthermore, a detailed study of the influence of different antenna geometries on the capacity of OFDM based spatial multiplexing systems appears to be of interest. This problem has been studied to some extent in [17] for the narrow-band frequency-flat fading case.

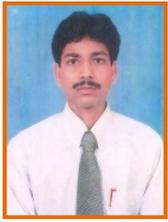

**Prof. Nirmalendu Bikas Sinha** received the B.Sc (Honours in Physics), B. Tech, M. Tech degrees in Radio-Physics and Electronics from Calcutta University, Calcutta,India,in1996,1999 and 2001, respectively. He is currently working towards the Ph.D degree in Electronics and Telecommunication Engineering at BESU. Since 2003, he has been associated with the College of Engineering and Management, Kolaghat. W.B, India where he is currently an Asst.Professor is with the department of Electronics & Communication Engineering & Electronics & Instrumentation Engineering. His current research Interests are in the area of signal processing for high-speed digital communications, signal detection, MIMO, multiuser communications,Microwave /Millimeter wave based Broadband Wireless Mobile Communication ,semiconductor Devices, Remote Sensing, Digital Radar, RCS Imaging, and Wireless 4G communication.  He has published large number of papers in different international Conference, proceedings and journals.He is presently the editor and reviewers in different international journals.

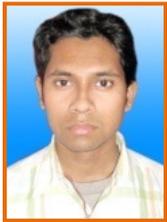

**Prosenjit Kumar Sutradhar** is  pursuing B.Tech in the Department of Electronics & Communication Engineering at College of Engineering and Management, Kolaghat, under WBUT in 2011, West Bengal, India. His areas of interest are in Microwave /Millimeter wave based Broadband Wireless Mobile Communication and digital electronics. He has published multiple publication in international journals.

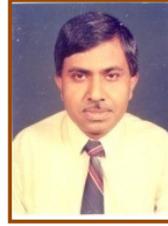

**Dr. Monojit Mitra** is an Assistant Professor in the Department of Electronics & Telecommunication Engineering of Bengal Engineering & Science University, Shibpur. He obtained his B.Tech, M.Tech & Ph. D .degrees from Calcutta University. His research areas are in the field of Microwave & Microelectronics, especially in the fabrication of high frequency solid state devices like IMPATT. He has published large number of papers in different national and international journals. He has handled sponsored research projects of DOE and DRDO. He is a member of IETE (I) and Institution of Engineers (I) society.